\documentclass[conference]{IEEEtran}

\usepackage{graphics}
\usepackage{graphicx}
\usepackage{subfigure}
\usepackage{amsthm}
\usepackage{color}
\usepackage{amssymb}
\usepackage{amsmath}

\newtheorem{definition}{Definition}

\begin{document}

%\title{How Nearsightedness Can Harm Performances\\ in Disruption-Tolerant Networking}
%\title{The Impact of Eagle-Eyedness on Disruption Tolerant Networking Performances}
%\title{Wait But Not That Much: Using Neighborhood Beyond One Hop in Disruption-Tolerant Networks}
\title{Using Neighborhood Beyond One Hop in Disruption-Tolerant Networks} %\\ \medskip \Large Should we wait longer or take a look around?}

\author{Tiphaine Phe-Neau$^{\dagger}$, Marcelo Dias de Amorim$^{\dagger}$, and Vania Conan$^{\ddagger}$\vspace*{1mm}\\ 
{\small
\begin{tabular}{c c}
  $^{\dagger}$CNRS and UPMC Sorbonne Universit{\'{e}}s & $^{\ddagger}$Thales Communications\\
  \{tiphaine.phe-neau, marcelo.amorim\}@lip6.fr & vania.conan@fr.thalesgroup.com \\
\end{tabular}}
}

\maketitle

\begin{abstract}
Most disruption-tolerant networking (DTN) protocols available in the literature have focused on mere contact and intercontact characteristics to make forwarding decisions. Nevertheless, there is a world behind contacts: just because one node is not in contact with some potential destination, it does not mean that this node is alone. There may be interesting \text{end-to-end} transmission opportunities through other nearby nodes. Existing protocols miss such possibilities by maintaining a simple \text{contact-based} view of the network. In this paper, we investigate how the vicinity of a node evolves through time and whether such information can be useful when routing data. We observe a clear tradeoff between routing performance and the cost for monitoring the neighborhood. Our analyses suggest that limiting a node's neighborhood view to three or four hops is more than enough to significantly improve forwarding efficiency without incurring prohibitive overhead. 
\end{abstract}

%\begin{IEEEkeywords}
%Disruption-Tolerant Networks, WAIT protocol.
%\end{IEEEkeywords}

\section{Introduction}
\label{sec:introduction}

As our urban society lives on, the more technologically nomadic its citizens get. During their daily commuting, people carry electronic devices like smartphones, portable game stations, or laptops. Such objects embed wireless interfaces and important storage abilities traveling with their owners. Noticing those advantages, an emerging networking paradigm adapted itself to leverage these opportunistic contacts, namely disruption-tolerant networking~\cite{fall.sigcomm03}. 

Disruption-tolerant networks lack overall network knowledge. DTN nodes are only aware of what they learned via encounters. To forward information, nodes try to make the most of their local knowledge. Attentive to these difficulties, researchers created ingenious forwarding techniques in DTNs. Some approaches rely on full flooding via encountered nodes like Epidemic forwarding but are quite wasteful in terms of energy and resources optimization~\cite{vahdat.epidemic}. Wiser epidemic protocols emerged as with PRoPHET or Spray-and-Wait, where nodes choose their next hops based on probabilistic likeliness of meeting the destination or through distributed flooding~\cite{lindgren.sigmobile03,spyropoulos.wdtn05}. 

Some other techniques choose to use the social behavior of the participants. As in a city people tend to cluster into communities around different points of interests, \"Ott et al. presented a protocol leveraging end-to-end and multihop DTN paths~\cite{ott.chants06}. End-to-end paths occur among connected components whereas DTN ones happen between these temporary components. Sarafijanovic-Djukic et al. made a similar observation in the VANET environment~\cite{djukic.secon06}. Later, Heimlicher and Salamatian laid their study over the groundwork that mobile wireless networks tend to have connected crowds~\cite{heimlicher.mobihoc10}. The main punch line for all these studies is: for each node, there are immediate neighborhood structures to leverage on. 

%Heimlicher et al. defended the existence and usage of partial paths aka temporal sequence of contemporaneous multi-hop path in DTN routing~\cite{heimlicher.infocom09}. 

The inherent nature of DTN pushes us to use local information to make routing decisions, but why do we restrain our knowledge to contacts information as featured in Epidemic, PRoPHET or Spray-and-Wait? A node's current connected component may help us take forwarding decisions. Here, we try to see beyond a node's mere contacts and dig deeper into its immediate neighborhood. 

To detect the impact of neighborhood beyond one hop in DTN protocols, we perform our analysis over the most basic DTN forwarding strategy, namely the WAIT protocol. In this protocol, the source stores the message until it meets the destination. The main criticism on the WAIT approach is that the source may wait for a quite long time before being able to deliver the message or, worst, to completely fail delivering it. In this study, we rely on our previous findings that nodes are frequently nearby even if not in direct contact and therefore, there are non negligible $n$-hop paths around nodes to use~\cite{pheneau.iscc11}. Localized neighborhood knowledge can be an important asset for DTN nodes. However, the main challenge in providing extended neighborhood knowledge to nodes is to find a good balance between efficiency and probing costs. In fact, we can wonder how far a node should probe its neighborhood for surrounding knowledge. The more information about the network, the better decisions we can make but the higher the costs induced. 

We first identify the costs of neighborhood knowledge through simple yet realistic overheads. Then we analyze the gains that neighborhood knowledge brings in the WAIT protocol and mitigate our observations according to the aforementioned overheads. Finally, we issue an empirical scope \text{limitation} for neighborhood monitoring that brings both improved efficiency and constrained costs. Performances comparison with other DTN protocols is beyond the scope of our study and will be considered as a further research topic.

As a summary, this paper makes the following key contributions:

\begin{itemize}

	\item 	We confirm the power of localized neighborhood knowledge in the design of efficient forwarding algorithms for DTNs.  
	
	\item		We define custom performance metrics for the WAIT protocol and quantify the performance gains brought by neighborhood knowledge.

%	\item 	We identify a tradeoff between network knowledge and resources overhead.

	\item 	We show an empirical limitation of neighborhood knowledge allowing better waiting times and constrained monitoring costs.

\end{itemize}

In the remainder of this paper, we first formulate the problem and the necessary background knowledge for its definition in Sections~\ref{sec:problem}, \ref{sec:ewait}, and~\ref{sec:cost}. Then, we evaluate this proposal on a number of synthetic and real-world mobility traces in Section~\ref{sec:evaluation}. We finally conclude the paper and suggest ideas for future work in Section~\ref{sec:conclusion}.

%As for now, DTN protocols only use the knowledge of contacts to make their decision concerning transmissions. For instance, the Spray and Wait routing protocol sends copies of a message $m$ to an encountered node E only if E does not have copies of $m$ yet~\cite{spyropoulos.wdtn05}. But what if E never has any transmission possibilities to $m$'s destination? And what if E also forwards to other nodes of this type? This ends up in a complete waste of resources. 

%Current DTN protocols make their forwarding decision following two predominant ideas: they either take random decision for their next hop nodes either choose them based on some likeliness of encounters schemes. We pursue a different path and defend the fact that an accurate neighborhood knowledge is enough to make a good relay nodes choice. 

%This type of knowledge has already been introduced under the name of \textit{$n$-ary intercontact} with the \textit{favorable intercontact} notion~\cite{pheneau.iscc11}. End-to-end path between a pair of nodes may indicate low delay transmission opportunities. Therefore, this connectivity knowledge can have major impact on protocol performances. 

\section{Problem Formulation}
\label{sec:problem}

In the current DTN landscape, protocols derive their transmission decision based on their contact observations. We consider that a contact happens between two nodes whenever they are within each other's communication range.

The performance of DTN forwarding protocols is related to two main parameters. The first is the \textit{time-to-live} (TTL), which is used to bound resource utilization or network availability through time; whenever the TTL expires, messages are dropped, which guarantees that messages do not travel indefinitely in the network. The second and most important parameter is the \textit{waiting time.} It is the duration a node waits before sending a message toward the destination. The lower the waiting time, the more chances we have to respect a message TTL and the better the induced delivery delay. In the case of the WAIT protocol, the waiting time only stops when the source meets the destination (no intermediate relays). Note that the two parameters are closely related. The waiting time has a particular meaning as it is related to the user experience: any user notices the duration before its message gets delivered (if it gets delivered at all) and judges a service accordingly.

\subsection{Why monitoring contacts only?}

It is easy to understand that at work, at school, or in a restaurant, a person $A$ may be collocated with many people but $A$ is not always or ever in direct contact with some of them. Imagine $A$ wants to send a message to $B$, which dwells $2$ hops away from $A$ for 30 minutes and then leaves the place. $A$'s waiting time would be infinite, as they never came into contact, but $A$ could have sent the message earlier when they were nearby. 

\begin{figure}[t]
	\begin{center}
	\includegraphics[width=\columnwidth]{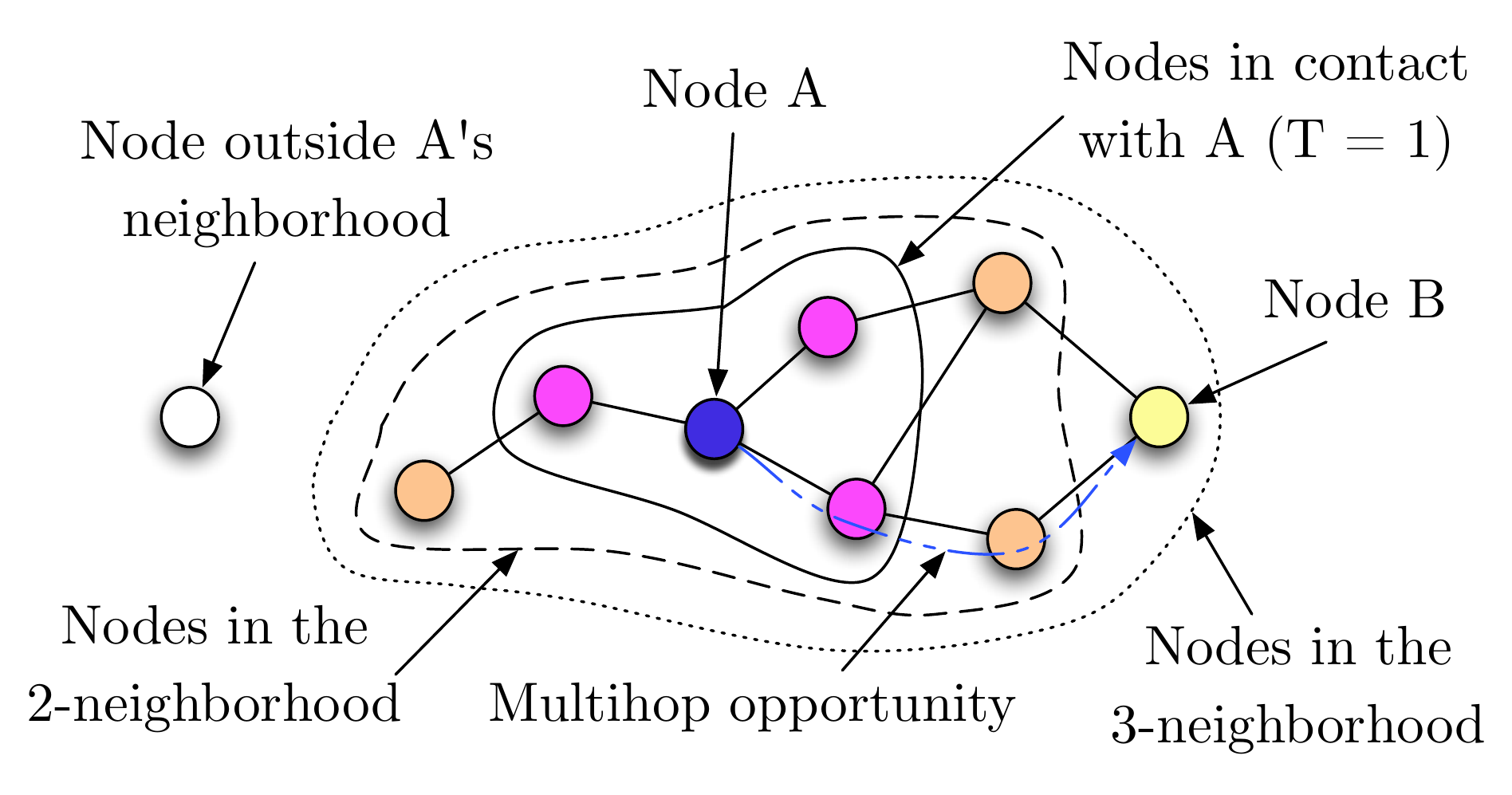}  
	\caption{Node $A$'s instantaneous $T$-neighborhood. When $T=1$, $A$ only knows nodes in contact with him. With $T=2$, $A$ know all nodes within a $2$ hop distance (his $2$-neighborhood) and so on. $A$ has an end-to-end path, of at most length $T$, to any member of his $T$-neighborhood. Note the $3$-hop paths from $A$ to $B$.}
	\label{fig:t_neighborhood}
	\end{center}
\end{figure}

\subsection{Why don't we take a look around?}

Consider now that, in the example above, $A$ knows the topology of the network. $A$ would actually know that the destination is within end-to-end reach and instead of waiting until it meets $B$, it could perform multihop communication and benefit from the $n$-hop opportunity. This is illustrated in Fig.~\ref{fig:t_neighborhood}. 

DTN protocols rely on contacts as they are easy to gather, while extended neighborhood knowledge is more costly. Due to the inbuilt nature of DTNs, offering nodes a consistent and full knowledge of the network topology is unrealistic. An alternative would be to make nodes have information on the connected component (CC) they are in. This would allow each node to know to whom it has a contemporaneous path. \text{However}, as nodes do not know a priori the size of the connected component, it becomes difficult to limit the control overhead. Still, by limiting the scope of a node vision (to nodes below $2$ or $3$ hop distance for instance), we also limit the overhead increase. This method raises an interesting question: \textit{How does neighborhood knowledge impact forwarding performance?}

To answer this question, we investigate the impact of the neighborhood knowledge in the WAIT protocol by focusing on the waiting parameter~-- the time a node waits before being able to send a message straight to the destination.
\section{The $n$-ary intercontact characterization}
\label{sec:ewait}

We rely on the DTN characterization based on \textit{contact} and \textit{$n$-ary intercontact} to discriminate a node's connected component~\cite{pheneau.iscc11}. For each node, any neighbor staying at a $1$-hop distance is considered to be in contact. In our definition, we say that other nodes are in \text{$n$-ary intercontact}. This intercontact characterization holds two main notions:

\begin{definition}
\textbf{Favorable intercontact.} Two nodes are considered in ``favorable'' intercontact with parameter $n$ when there is a contemporaneous shortest path of length $n\in[2; \infty[$ separating the two nodes under consideration. These nodes belong to the same connected component.
\end{definition}

Note that the parameter $n$ helps represent the limit threshold we want to observe in this study. 

\begin{definition}
\textbf{Pathless intercontact.} In opposition to favorable situations, ``pathless'' intercontact indicates the lack of \text{end-to-end} paths between the two nodes, i.e., $n=\infty$. 
\end{definition}

For further details on the $n$-ary intercontact characterization, please refer to~\cite{pheneau.iscc11}. Using this intercontact characterization, we introduce a threshold $T$ between the favorable and pathless intercontacts to limit the knowledge a node will have about its neighborhood. All nodes in favorable intercontact of parameter $n \le T$ are in the node connected component. Otherwise, they are considered outside the node $T$-neighborhood. The additional favorable intercontact knowledge allows nodes to know members of their $T$-neighborhood, i.e., all nodes to which they have an end-to-end path of length $l \leq T$. By choosing $T = 1$, we define that a node only knows its contacts. With $T = 2$, it knows all its contacts with his 2-hop neighbors and so on. Note that to be sure to get complete knowledge of the connected component, we must set $T = \infty$. In Fig.~\ref{fig:t_neighborhood}, we show a node's knowledge depending on the $T$ threshold. 

Now, whenever node $A$ wants to send a message to node $B$, $A$ scans its neighborhood up to $T$ hops. If $B$ arrives within $A$'s $T$-neighborhood, $A$ sends the message to $B$ via a multihop contemporaneous path. To analyze raw performance of waiting times between the mere WAIT protocol and the WAIT protocol with neighborhood awareness, we do not consider any specific message sizes or throughputs that would impact delivery times. We focus on waiting times and on additional data impacting the control overhead.

\section{Cost analysis}
\label{sec:cost}

To take into account the costs of multihop messaging and neighborhood monitoring, we identified two main sources of overhead. We use the message as the unit of comparison.

\subsubsection{Data Overhead ($D_{o}$)} represents the total cost to deliver a message. Clearly, any protocol with extended neighborhood knowledge is costlier than its simple version. Whenever the source switches to multihop transmission mode, the message follows a contemporaneous end-to-end path to the destination and has to sustain several store and forward processes. The ``extra'' cost of such a communication, in terms of additional messages sent, is the number of hops between $A$ and $B$ minus one:

\begin{equation}
D_{o} = n - 1.
\end{equation}

\subsubsection{Neighborhood Knowledge Overhead ($N_{o}$)} 
\begin{figure}[t]
	\begin{center}
	\includegraphics[width=\columnwidth]{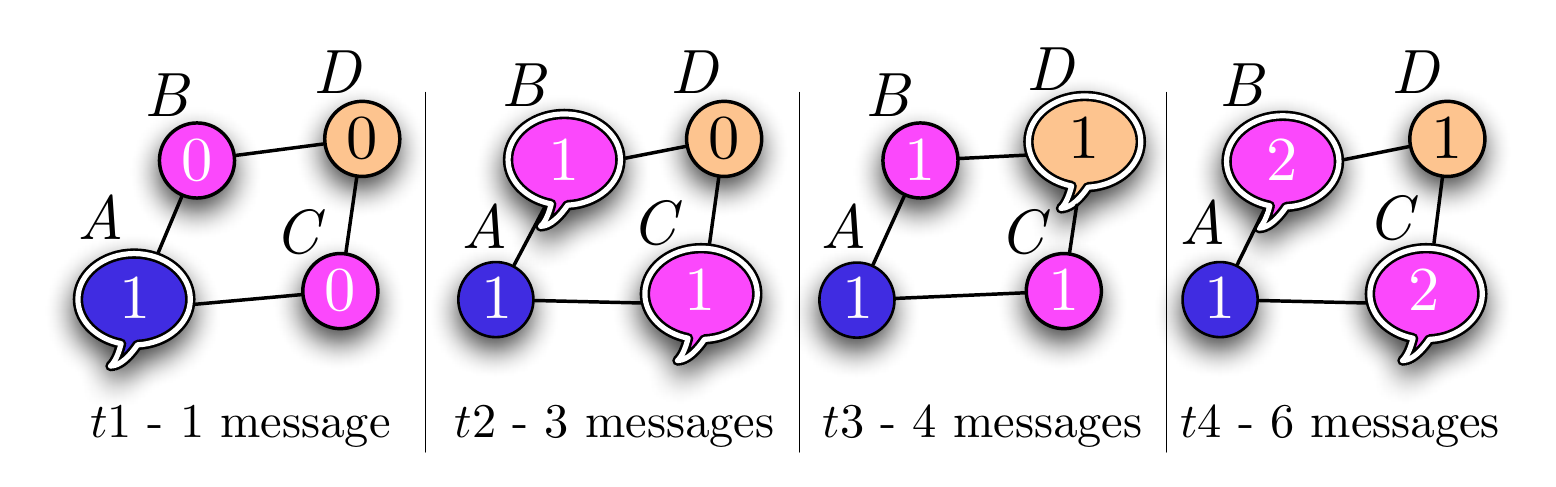} 
	\caption{Neighborhood knowledge discovery technique. At $t1$, $A$ ignites the discovery by broadcasting a message with a TTL set to $2$. His contacts, $B$ and $C$ receive the message. At $t2$, they broadcast a message with a TTL set to $2-1=1$. At $t3$, $D$ received discovery messages with a TTL of $1$, then broadcasts his reply. At $t4$, $B$ and $C$ aggregate all replies they received and send their knowledge to $A$. In the end, we obtain 6 sent messages.}
	\label{fig:nko_technique}
	\end{center}
\end{figure}

represents the overhead to gather information about the neighborhood. To get all nodes $T$-neighborhood, the basic approach consists in sending epidemic probes with an upper threshold of $T$. Node $A$ broadcasts a discovery message (${\tt DM}$) to its contacts with a TTL set to $T$. All nodes who received the ${\tt DM}$ rebroadcast this message with a TTL set to $T-1$, and so on. We assume that each transmission is acknowledged (see Fig.~\ref{fig:nko_technique} for a detailed example). This leads to a cost of:

\begin{equation}
N_{o} = 2\times \text{card}(\text{CC of size} < T) + \text{card}(\text{CC of size} = T) + 1,
\end{equation}

\noindent where ``card'' stands for cardinality and ``CC'' for connected component. $N_{o}$ does not depend on the path length that ${\tt DM}$s have to cross. With little aggregation, $N_{o}$ only depends on the number of neighbors in a node's connected component. $N_{o}$ is responsible for most of the overhead in our analysis as it consists in frequent neighborhood monitoring.

\section{Evaluation}
\label{sec:evaluation}

To quantify the performance gains enabled by neighborhood awareness, we simulate the WAIT protocol with and without $T$-neighborhood ($T \ge 2$) knowledge using real-world mobility data and the ONE simulator~\cite{keranen.simutools09}.

\subsection{Datasets}

To grasp our issue in different situations, we confront both scenarios to two synthetic mobility models and two realistic datasets. The synthetic models are:

\begin{itemize}

	\item \textbf{RandomTrip} is a well-known mobility model making up known issues of Random Waypoint~\cite{palchaudhuri.anss05}. We simulated 20 nodes during 9 hours with speeds ranging from 0 to 7~m/s on a 50$\times$60~m$^2$ surface. This scenarios intends to emulate a working day structure.

	\item \textbf{Community} is a mobility model founded on social network theory~\cite{musolesi.comcom07}. This model emphasizes the human tendency to aggregate in a societal way~-- based on relationships. We simulated 50 nodes during 9 hours on a 1,500$\times$2,500~m$^2$ surface. This dataset denotes an urban-wise environment. 

\end{itemize}

We also consider two real-world datasets probing contacts between nodes via motes. Motes log the presence of other motes within a 10-meter wireless range.

\begin{itemize}

	\item \textbf{Infocom05} retraces contacts between 41 attendees during a 5-day conference~\cite{chaintreau.tmc07}. We focus on a 12-hour period of the second day which presents the highest network activity. Each device performs a scan every 120 seconds.

	\item \textbf{Rollernet} occurs during a 3-hour rollerblading tour in Paris~\cite{tournoux.infocom09}. This trace has a finer beaconing granularity of 15 seconds and represents a highly dynamic sport event.

\end{itemize}

We investigate in the following how much we can reduce the waiting times using neighborhood awareness.

\subsection{Delay observations}

For each mobility trace and each pair of nodes, we randomly generated 10 messages at different time instants. We chose to generate sparse messages for waiting times to better reflect the impact of neighborhood monitoring. The most symptomatic situation arises when a pair of nodes never come in contact, but once and a while they belong to the same connected component. In this situation, the WAIT protocol drops the message whereas the neighborhood-aware variant can manage to forward it correctly. 

As scarce as this situation may sound, it happens for 10\% of pair of nodes in \textit{Infocom05}, 53\% in \textit{Community}, and around 55\% of \textit{Rollernet} nodes. If these nodes try to send a message using the WAIT protocol, they will fail. These fractions of nodes have \textbf{infinite waiting delays} when WAIT is in use. Otherwise, with the neighborhood-friendly version, they manage to deliver messages with bounded waiting times.

\begin{figure}[t]
	\begin{center}
	\includegraphics[width=\columnwidth]{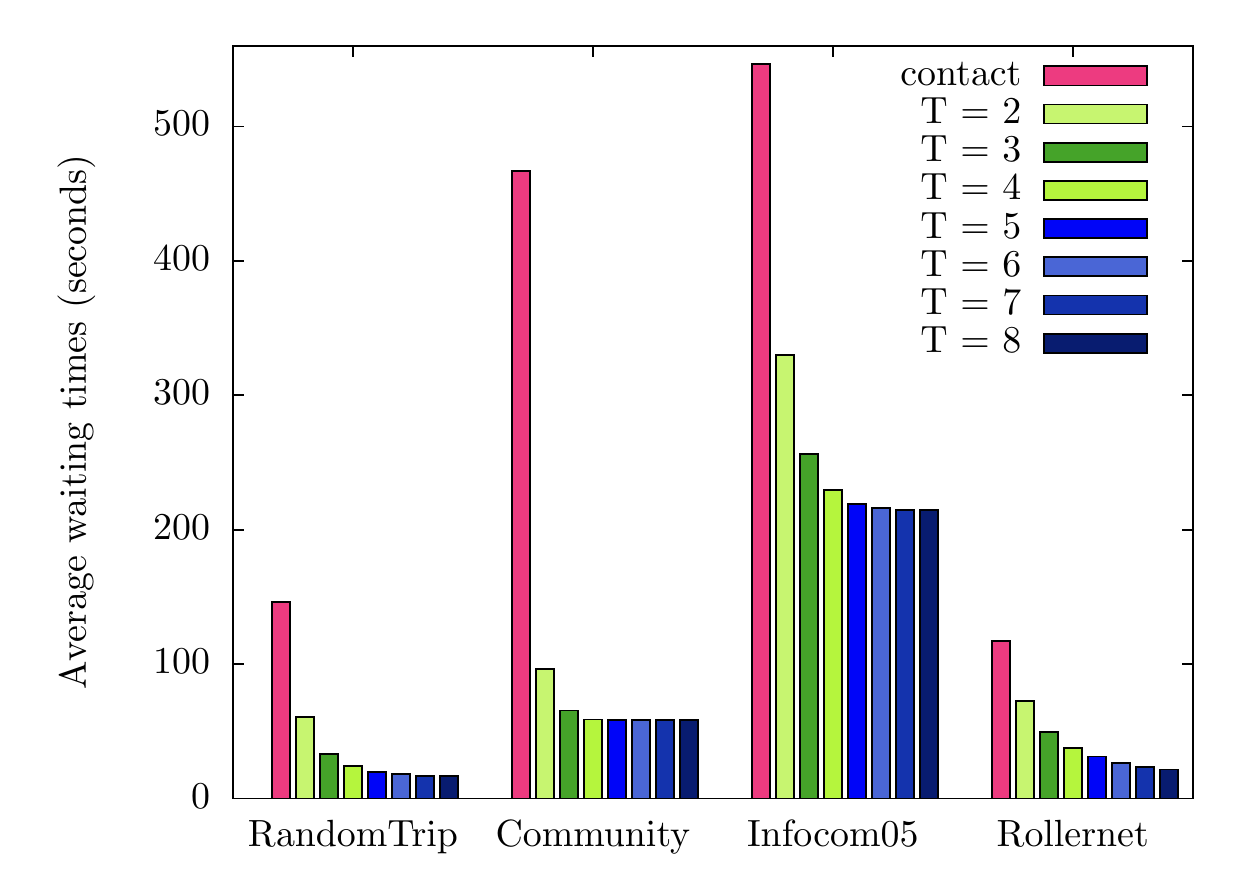} 
	\caption{For each dataset, we represent averaged waiting times according to the threshold $T$. For all traces, there is a clear improvement between the first and second bar (contact only vs. $2$-neighborhood). Being aware of a node's $T$-neighborhood can lead to divide waiting times by 4 like in \textit{Community}. The higher the $T$, the better the waiting delays, yet, above $T > 4$, gains become negligible.}
	\label{fig:wtimes}
	\end{center}
\end{figure}

For these nodes with \textbf{bounded waiting delays}, we analyze to which extent neighborhood knowledge helps lower their waiting times. In Fig.~\ref{fig:wtimes}, we show the averaged pairwise waiting times for each dataset. Each bar represents the average waiting delay we obtain with $T$-neighborhood probing. For every dataset, between the first and second bars, we notice significant reduction in the waiting times: 40\% in \textit{Infocom05} and \textit{Rollernet}, 57\% in \textit{RandomTrip}, and around 80\% in \textit{Community}. We observe that, although we keep reducing the waiting delays, the gains for $T > 4$ are much smaller. This corroborates our first feeling that localized knowledge should be enough and suggests that we can, in practice, keep $T$ small (two or three).

On the one hand, when nodes have no contact opportunities and therefore an infinite waiting delay, neighborhood monitoring can bind waiting times and enable forwarding possibilities. On the other hand, when nodes do have contact opportunities, neighborhood monitoring still helps in lowering waiting delays (up to a factor of four in our scenarios). $T$-neighborhood monitoring clearly helps reducing waiting delays; but it also ignites some costs, which we investigate in the next subsection. 

\subsection{Overhead study}

Supporting neighborhood knowledge monitoring does not come for free. Any node needs to probe its neighborhood and create a flow of messages around. 

\subsubsection{Impact of neighborhood knowledge overhead}

There are many strategies for connected component gathering, from link state-like solutions to flooding techniques. For our study, we chose to compare two naive behaviors:

\begin{itemize}

	\item Nodes keep monitoring their $T$-neighborhood at regular time intervals (called ${\tt CS}$ hereafter).

	\item Nodes monitor their $T$-neighborhood when they have a message to send and stop when the message TTL expires (called ${\tt TS}$ hereafter).

\end{itemize}

\begin{figure}[t]
	\begin{center}
	\includegraphics[width=\columnwidth]{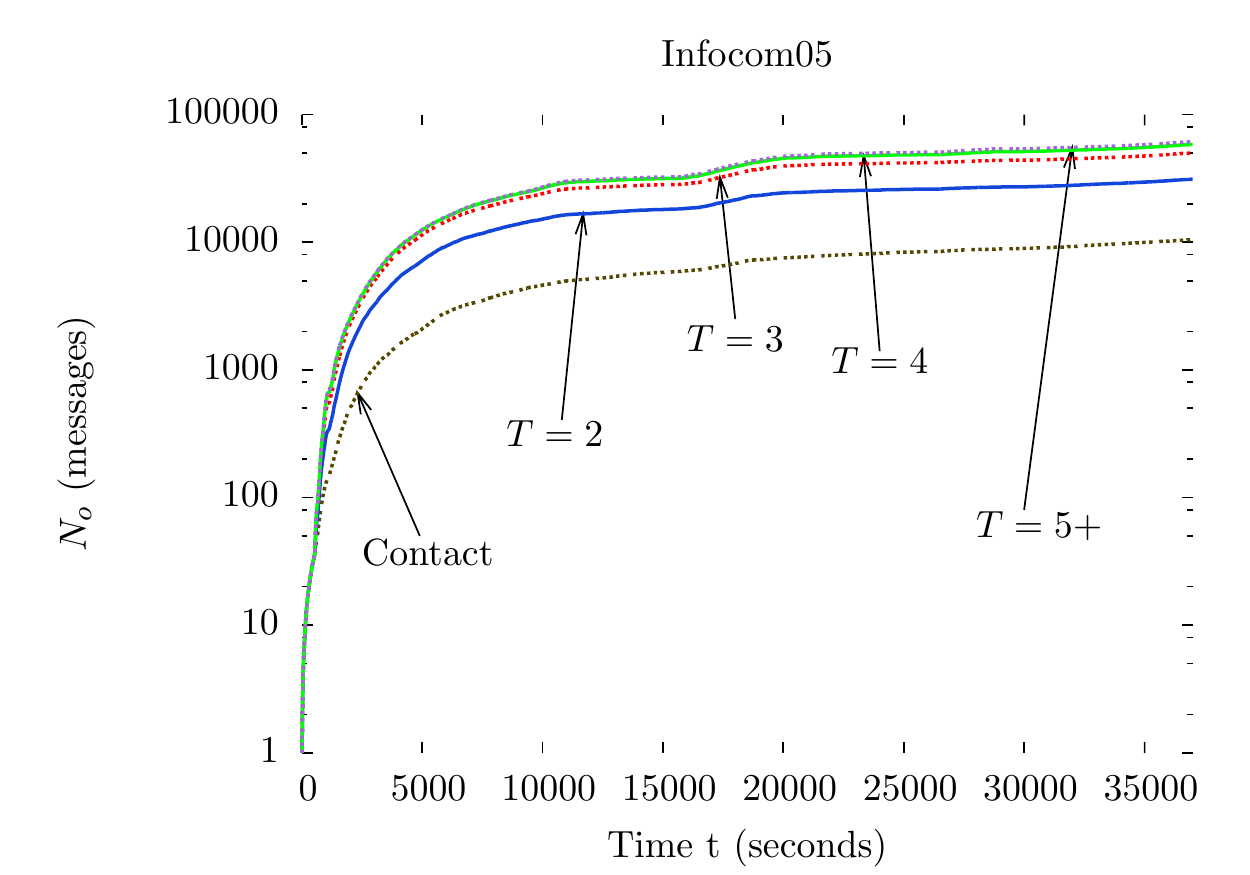} 
	\caption{Neighborhood Knowledge Overhead ($N_{o}$) in terms of message sent by the discovery technique ${\tt CS}$ for a node in the \textit{Infocom05} dataset. On average, probing $T$-neighborhood with $T>4$ costs as much as probing the \text{$4$-neighborhood}. This version of neighborhood probing is very expensive. Note the logscale on the $y$-axis.}
	\label{fig:nko_full}
	\end{center}
\end{figure}

In Fig.~\ref{fig:nko_full}, we represent $N_{o}$ during the whole experiment duration for a node in the \textit{Infocom05} with ${\tt CS}$ probing every 30 seconds. Each curve represents the behavior for each threshold $T$. There are major differences between thresholds. Monitoring only contacts induces fewer overhead than any deeper neighborhood monitoring. Beyond $T=4$, there are no noticeable differences for $N_{o}$. Overall behaviors are quite alike and depend on the surrounding density.

\begin{figure}[t]
	\begin{center}
	\includegraphics[width=\columnwidth]{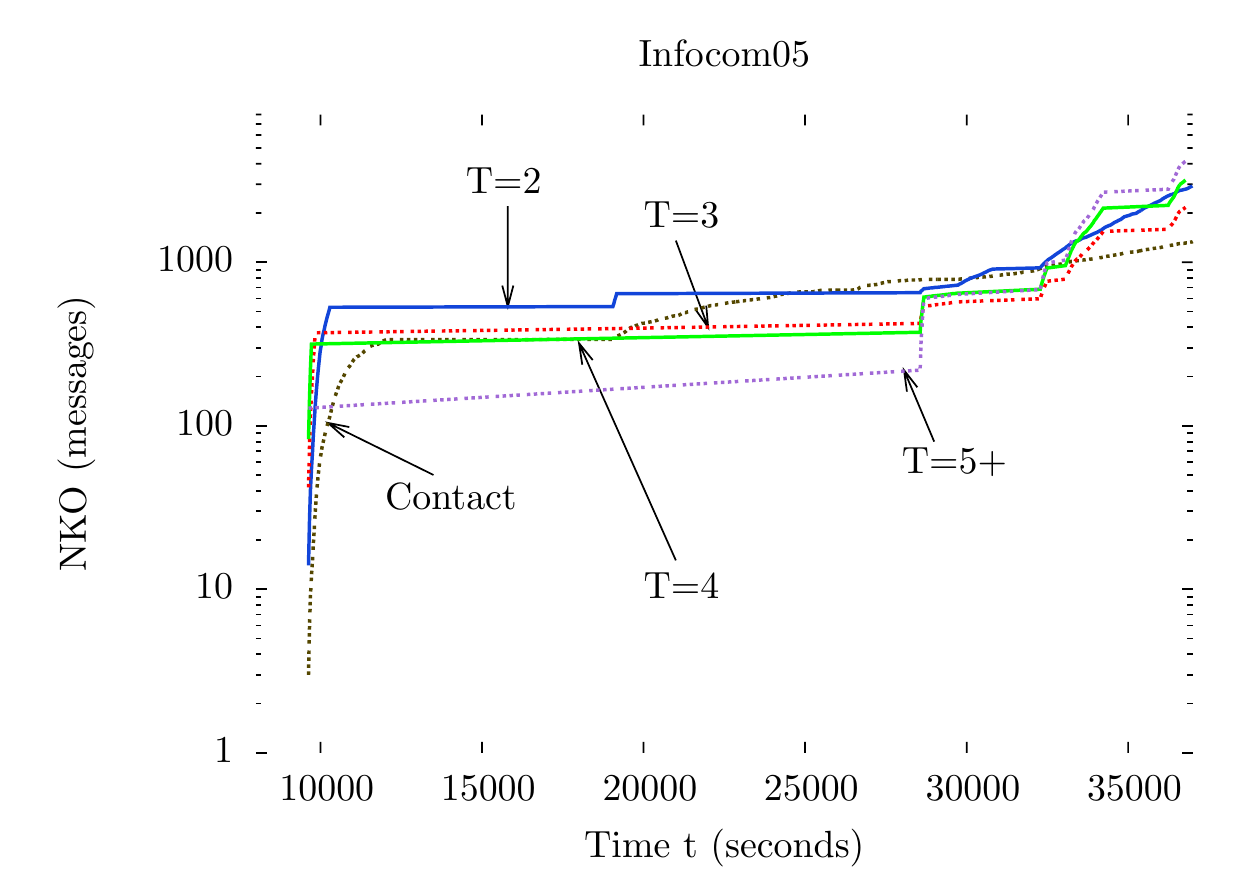} 
	\caption{Neighborhood Knowledge Overhead ($N_{o}$) in terms of message sent by the discovery technique ${\tt TS}$ for a pair of node in the \textit{Infocom05} dataset. Contact monitoring drops $9/10^{th}$ of messages in this situation and keeps monitoring its contacts without being able to deliver any messages. An interesting observation is how, for the same number of delivered message, sensing \text{$3$-neighborhood} (or beyond) ends up cheaper than observing \text{$2$-neighborhood}. Here, $T \geq3$ leads to shorter waiting delays and shorter probing periods than with $2$-neighborhood. Note the logscale on the $y$-axis.}
	\label{fig:nko_wtimes}
	\end{center}
\end{figure}

In Fig.~\ref{fig:nko_wtimes}, we plot $N_{o}$ of the same source node as before. This time, we use the ${\tt TS}$ method for neighborhood analysis. The reason we have noticeable jumps in all curves is, when the destination comes into the source's $T$-neighborhood, this latter stops monitoring its surroundings. Contact monitoring drops all but one message and is only plotted for the reader's information. An interesting result is how, for the same number of delivered messages ($7$ messages), probing the \text{$3$-neighborhood} and beyond gives better results than probing only the \text{$2$-neighborhood} in terms of $N_{o}$. The reason is that the faster the source finds the destination, the shorter the waiting delay and the lower the $N_{o}$. 

For the tradeoff analysis on lowering waiting delays and overheads costs, we will consider ${\tt TS}$ as the neighborhood monitoring strategy. For WAIT, a constant monitoring ${\tt CS}$ is a wasteful strategy in terms of messages or energy use.

\subsubsection{Impact of data overhead}

$N_{o}$ seems to be the most expensive in terms of messages sent, yet, we also have to take into account the $D_{o}$: the amount of messages generated over an end-to-end path transmission. $D_{o}$ adds an insignificant amount of messages to $N_{o}$. It is important to underline that having a large $D_{o}$ (i.e., a long path between the sender and the destination) can lead to undelivered messages. This is why one would prefer shorter $T$.

\subsection{Tradeoff analysis}

We have shown how using neighborhood monitoring can reduce waiting delays. In Fig.~\ref{fig:wtimes}, we observe that a larger $T$ leads to lower waiting delays. Yet, neighborhood monitoring is an expensive process. In Fig.~\ref{fig:nko_wtimes}, we see how $N_{o}$ evolves with time. With a simple probing technique, we manage to constrain message overheads and deliver a higher message rate than with the WAIT protocol.

For each node, we analyze the average number of neighbors in their $T$-neighborhood. TABLE~\ref{tab:avg_cc} shows this value for the whole dataset duration. We understand that above a certain threshold $T_t$, a node's $T$-neighborhood does not expend much (except for the \textit{RandomTrip} dataset, which has a random movement pattern and a high density). In \textit{Community} or \text{\textit{Infocom05}}, a node's $T$-neighborhood does not grow significantly anymore above $T_t=4$. %In average, monitoring the $T$-neighborhood with $T > 4$ costs as much as monitoring the $4$-neighborhood should it be with both probing techniques. As a result, one would conclude that setting up a $4$-neighborhood (or more) monitoring for each node would be interesting.

%Because of the people-driven nature of these DTNs, we can lower the threshold for neighborhood monitoring. 

\smallskip\noindent\textbf{Arguments.} People do move around with their devices. Each of them may recompute its neighborhood every once and a while after a journey. With the whole network movement, each node is more likely to find its destination even with a shorter threshold. We have seen how observing a node's $3$-neighborhood is interesting enough to lower the waiting delays and constrain messages overheads. Monitoring $3$-neighborhood can even reduce overheads by lowering such a delay compared to sensing $2$-neighborhood. Moreover, lower $T$ thresholds enable lower $D_{o}$ and promote quicker message forwarding. 

\begin{table}[t]\small%\small
\caption{\label{tab:avg_cc}Average number of neighbors in a node's $T$-neighborhood (whole dataset duration).}
\begin{tabular}{c|c|c|c|c|c|c|c|c}
 & \multicolumn{8}{c}{$T$} \\
\textbf{Dataset}                  & $1$	& $2$ & $3$	& $4$	& $5$	& $6$ & $7$	&$8+$	  \\ \hline
\textit{Community}	& 2.0 	& 4.0 	& 4.6 	& \textbf{4.7}	& 4.7 	& 4.7 	& 4.7 	& 4.7	  \\ \hline
\textit{RandomTrip}	& 2.0 	& 3.2 	& 4.7 	& \textbf{5.7} 	& 6.3 	& 6.7 	& 6.9 	& 7.1	  \\ \hline
\textit{Infocom05}	& 1.5 	& 3.8 	& 5.3 	& \textbf{6.0} 	& 6.4 	& 6.4 	& 6.4 	& 6.4	  \\ \hline
\textit{Rollernet}		& 1.4	 & 3.2	& 4.7 	& \textbf{5.7} 	& 6.3 	& 6.7 	& 6.9 	& 7.0 	  \\
\end{tabular}
\end{table}
\section{Summary and outlook}
\label{sec:conclusion}

In this paper, we examine the impact of neighborhood awareness on the waiting time in a variation of the simple WAIT protocol. Most DTN techniques only focus on sensing direct contacts and do not inquire about their nearby \text{neighbors}. This strategy is the most straightforward and rational. However, we cannot deny the sociological nature of DTNs. People do not wander randomly in a city. They gather around specific persons or locations. To our opinion, ignoring a node's immediate neighborhood results in a loss of useful information.

Our findings show that neighborhood probing significantly improves the WAIT protocol in terms of waiting delays. When delays used to be infinite, they are now bounded. When delays were high, they are now lowered by a factor up to $4$ in our scenarios. Yet, neighborhood monitoring ignites messaging overhead. But, by limiting a node's neighborhood vision to a threshold $T$ of three or four, we constrain costs and still enhance performances. According to our results on the raw WAIT protocol, we encourage our community to consider potential performance gains that neighborhood knowledge could bring to more elaborated DTN schemes.

\section*{Acknowledgment}

This work is partially supported by the ANR project CROWD under contract ANR-08-VERS-006. 

\bibliographystyle{IEEEtran}
\bibliography{dtns}

\end{document}